\documentclass[aps,12pt,onecolumn,showpacs,floats,letterpaper]{revtex4}

\usepackage{times}
\usepackage{mathptmx}
\usepackage{graphicx}
\usepackage{dcolumn}
\usepackage{bm}
\bibpunct{}{}{,}{s}{}{}
\begin{document}
\preprint{}

\title{Electrically-detected magnetic resonance in ion-implanted Si:P nanostructures}

\author{D. R. McCamey $^{a,b)}$\email{dane.mccamey@unsw.edu.au}
, H. Huebl $^{c)}$, M. S. Brandt $^{c)}$, W. D. Hutchison
$^{a,d)}$,\\
J. C. McCallum $^{a,e)}$, R. G. Clark $^{a,b)}$ and A. R. Hamilton
$^{b)}$\\}

\address{ $^{a)}$Australian Research Council Centre of
Excellence for Quantum Computer Technology }

\address{ $^{b)}$School of Physics, The University of New
South Wales, Sydney, NSW 2052, Australia }

\address{ $^{c)}$Walter Schottky Institut, Technische
Universit\"{a}t M\"{u}nchen, Am Coulombwall 3, D-85748 Garching,
Germany}

\address{ $^{d)}$School of Physical, Environmental and
Mathematical Sciences, The University of New South Wales, ADFA,
Canberra, ACT 2600, Australia  }

\address{ $^{e)}$School of Physics, University of Melbourne,
VIC 3010, Australia}

\date{\today}


\begin{abstract}
We present the results of electrically-detected magnetic resonance
(EDMR) experiments on ion-implanted Si:P nanostructures at 5~K,
consisting of high-dose implanted metallic leads with a square
gap, in which Phosphorus is implanted at a non-metallic dose
corresponding to $10^{17}~\rm{cm^{-3}}$. By restricting this
secondary implant to a 100~nm~$\times$~100~nm region, the EDMR
signal from less than 100 donors is detected. This technique
provides a pathway to the study of single donor spins in
semiconductors, which is relevant to a number of proposals for
quantum information processing.

\end{abstract}


%


\maketitle


The ability to spectroscopically study the spin properties and
interactions of a small number of donors in semiconductors has
many applications such as the storage of classical information in
nuclear or electronic spins,\cite{recher} and is relevant to a
number of proposals\cite{vrijen, schenk} related to quantum
information processing (QIP). In particular, the construction of
QIP hardware utilizing Si:P has been discussed by Kane \cite{kane}
and Hollenberg.\cite{holl} In this context, the ultimate task is
the detection of the electron or nuclear spin state of single P
donors.

The detection of the spin resonance of donors in semiconductors
via electron spin resonance (ESR) is well established.\cite{feher}
However, the sensitivity of conventional ESR (where magnetisation
is measured) is limited to samples containing $10^{10}$~donors or
more.\cite{maier} This problem can be overcome by detecting
magnetic resonance via the effects of spin selection rules on
other observables, such as magnetic force \cite{rugar04},
radiative transitions\cite{Jelezko04}, or charge
transport.\cite{Elzerman04,Xiao04,Brandt}

Electrically detected magnetic resonance (EDMR), where a resonant
change of the dc conductivity is monitored\cite{Brandt}, was first
demonstrated on Si:P by Schmidt and Solomon.\cite{schmidt}
Subsequent studies of P in crystalline Si using EDMR were
performed both at very high \cite{honig} and very low magnetic
fields,\cite{stich95} as well as for P in amorphous\cite{brandt91}
and microcrystalline\cite{kanschat00} silicon. Electrical
detection of electron-nucleon double resonance
(EDENDOR)\cite{stich96} has been successfully demonstrated, also
on Si:P. As shown unambiguously by corresponding optically
detected magnetic resonance (ODMR) experiments on molecules, the
detection of ESR via spin-dependent electronic transitions can be
extremely sensitive, ultimately allowing the study of single
spins.\cite{koehler93, wachtrup93} Kawachi and
coworkers\cite{kawachi97} were able to observe EDMR from about
$10^4$ dangling bond defects in micron-size amorphous silicon
thin-film transistors. Stich et al.\cite{stich96} reported
successful detection of P in Si via EDMR in samples containing as
few as $10^6$ donors. However, no systematic study into the
sensitivity reachable with conventional EDMR experiments on donor
states has been published so far, in particular achieving even
better detection limits on P-doped silicon. Here, we present the
results of a systematic EDMR study of the sensitivity of
ion-implanted Si:P on samples where the smallest number of donors
in the active area is less than 100.

\begin{figure}[h!]
\includegraphics[width=6cm]{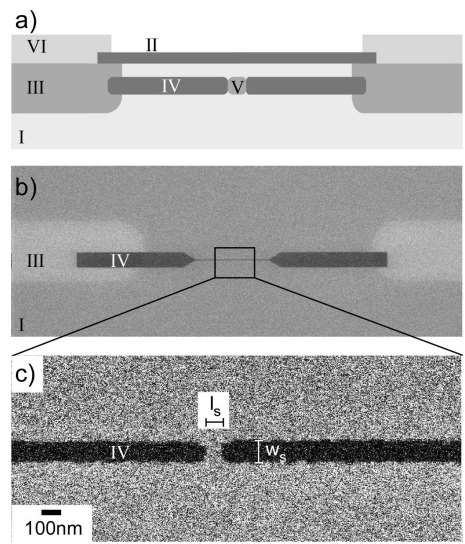} \caption{\label{fig:one} (color online) a)
Schematic view of the device showing the ohmic indiffusion (III),
the implanted metallic leads (IV) and the cluster implant location
(V). The other features are the substrate (I), the thin oxide (II)
and the Al/Au contacts (VI). b) SEM image of the the implanted
leads before RTA and c) an SEM image of the cluster region}
\end{figure}

The devices used for this study were fabricated on high
resistivity ($>$~8~k$\Omega$cm, corresponding to  $<10^{12}$
donors cm$^{-3}$) n-type silicon wafers (Fig. \ref{fig:one},
label I). First, ohmic contacts for the source and drain leads of
the device were defined via phosphorus indiffusion (III). A 5~nm
gate oxide (II) was then grown using a wet oxidation process.
TiAu (15~nm Ti,65~nm Au) markers, 100~nm~$\times$~100~nm in
dimension, were defined by electron-beam lithography (EBL) and
used to align subsequent EBL steps with an accuracy of $\pm$50~nm.
A 150~nm thick poly-methyl-methacrylate (PMMA) resist was applied
and patterned by EBL for use as a mask for ion-implantation of the
leads of width $w_s$ (labelled IV in Fig.~\ref{fig:one}) with P
implantation at an ion energy of 14~keV to an areal dose of
$\sim~1\times$10$^{14}$ cm$^{-2}$, corresponding to a doping
density of $\sim~4.0~\times~10^{19}$~cm$^{-3}$, well above the
Mott or metal-insulator transition at
$3.5\times10^{18}~\rm{cm^{-3}}$. The leads were doped to this
density as it has been shown \cite{murakami} that EDMR of highly
doped implanted Si:P does not show any hyperfine split resonance
signal. The mean implantation depth is $\sim$20~nm at this ion
energy. The tip of the leads define an active region (labeled V in
Fig. \ref{fig:one}), with a distance $l_s$ between the leads.


Using this basic contact geometry, three different devices were
fabricated. Type~1: Devices with a secondary P implant covering
the whole wafer. This implant was at an energy of 14~keV, with an
areal density of 5$\times$10$^{11}$ cm$^{-2}$, giving a maximum
doping density of 2$\times$10$^{17}$cm$^{-3}$. Type~2: Devices
where this secondary implant is limited to the area between the
leads (V) using a PMMA mask. The purpose of this type of sample is
to allow quantification of the number of donors being
investigated. Type~3: Control samples without secondary implant.
Following implantation, a rapid thermal anneal (RTA) at
1000~$^\circ$C for 5 seconds was performed to activate the donors
and repair the damage due to implantation\cite{mcc}. Afterwards,
ohmic contacts are formed by removal of the oxide in the contact
region, followed by deposition of Al(80nm)/Au(20nm) metallic
contacts.


EDMR measurements were performed in a modified Bruker ESR
measurement setup. The sample was illuminated with white light
from a halogen lamp, and a DC voltage V$_{DC}$ was applied to one
contact (source). The other contact (drain) was connected directly
to a current amplifier, and the output of the amplifier was fed
into a lock-in amplifier. Microwave radiation at a fixed frequency
$f_{\mu}$ in the X-band and a power of 50~mW was applied to the
circular dielectric cavity. Higher microwave power led to
significant microwave-induced currents, most probably due to
rectification by asymmetric contacts. The magnetic field was
modulated at $f_{\textrm{mod}}=1.234$~kHz with an amplitude of
0.3~mT. The external magnetic field $B$ was swept over 10~mT in
200 seconds. All spectra shown are corrected to a fixed microwave
frequency $f_{\mu}$=9.7~GHz. The field sweep was repeated and the
lock-in output averaged to obtain a high signal-to-noise ratio.
All measurements reported here were taken at $T=5$~K.


\begin{figure}[h!]
\includegraphics[width=6cm]{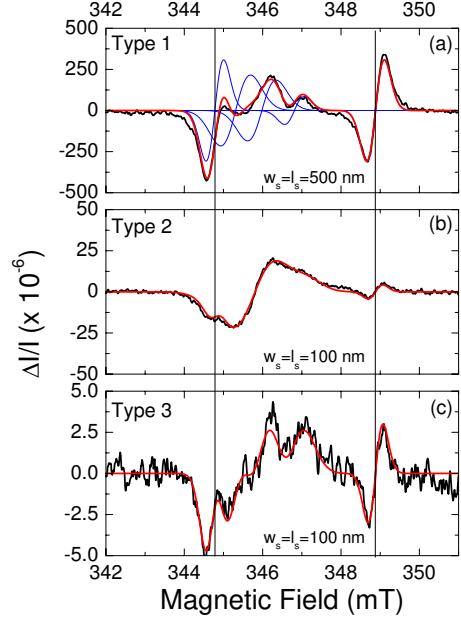} \caption{\label{fig:two}
(color online) EDMR signal $\Delta I/I$ vs. magnetic field for (a)
a Type 1 sample with $w_{s}=l_{s}=500$~nm, and (b) a Type 2 and
(c) Type 3 device both with $w_{s}=l_{s}=100$~nm. The red lines
are fits to the experiment. The blue lines in (a) show the
constituent lines of the fit. }
\end{figure}

Figure \ref{fig:two}~a shows the relative change $\Delta I/I$ of
the current vs. the external magnetic field for a Type 1 sample
with $w_s=l_s=500$~nm. The magnetic field positions of the
resonances were obtained by fitting the measured data with five
gaussian derivatives, as seen in Fig. \ref{fig:two}~a. The two
resonances due to the hyperfine splitting of P are clearly visible
with the expected 4.2~mT splitting.\cite{feher55} There is also a
smaller resonance observable at $B=346.81$~mT corresponding to
$g=1.9983$, which we attribute to exchange-coupled pairs of P
donors.\cite{feher55, cullis75} Additionally, two resonance lines
with $g$-factors of $g\approx 2.0031$ and $g\approx 2.0070$ are
observed for the orientation of the sample with $B\parallel
[110]$. Taking into account the limited resolution at X-band
frequencies and the power-induced broadening, these $g$-factors
are compatible with $g_{\perp}=2.0081$ and $g_{\parallel}=2.00185$
of the $P_{b0}$ defect\cite{stesmans98} and therefore are
attributed to this defect at the Si/SiO$_2$ interface.


Figure \ref{fig:two}~b shows the relative current change for a
Type~2 device with $w_{s}=l_{s}=100$~nm. The number of P implanted
into the active region is 50$\pm$8, as determined from the implant
parameters. Even with such a small number of phosphorus donors in
the active region, an EDMR signal intensity given by the
peak-to-peak current change of the high-field hyperfine-split P
resonance of $\Delta I/I=1.5\times 10^{-5}$ is easily detectable.
The defect signal observed in the Type~2 samples is dominated by a
single resonance at $g=2.005$, characteristic for the so-called
dangling bond signal also observed at Si/SiO$_2$
interfaces.\cite{cantin02}

Due to the use of P in the lead fabrication, the effect of the
straggle of the lead implantation in contributing donors to the
active sample area must be taken into account.
Figure~\ref{fig:two}~c shows the EDMR signal from a Type~3 device,
also with $w_{s}=l_{s}=100$~nm, where no P is implanted directly
into the gap area from a second implant. Also here, the
characteristic signature of hyperfine-split P is observable,
however with a smaller $\Delta I/I=6\times 10^{-6}$. This signal
is due to the straggle of the leads, which we now consider in
detail.

Spin-dependent hopping in Si:P at very high P concentrations has
been studied in detail, including samples where P incorporation
was obtained by implantation.\cite{murakami, murakami77} An
increase of the dark conductivity was observed under ESR
conditions at the central, exchange coupled resonance with
$g\approx 1.9985$\cite{young97} for samples above the Mott
transition, with typical values for $|\Delta I/I|$ decreasing from
$10^{-5}-10^{-7}$ at [P]$\simeq 6\times 10^{18}$~cm$^{-3}$ to
$10^{-10}$ at [P]~$>~3\times 10^{19}$~cm$^{-3}$.\cite{murakami,
kishimoto77, morigaki72} Therefore, a contribution from the leads
in our experiments will be limited to the central line at
$g=1.9985$ due to the high doping concentration. In contrast,
hyperfine-split lines can only arise from P in a local
concentration below $10^{18}~\rm{cm^{-3}}$.\cite{cullis75} The
amount of P with [P]$\leq 10^{18}~\rm{cm^{-3}}$ in a Type~3 device
with $w_s=l_s=100$~nm in area (V) between the leads and due to
straggle can be estimated from SRIM simulations\cite{srim} for our
implantation to be about $50\pm8$ considering the surface of the
leads facing area (V). Finally, due to the overlap of the two
implantation processes a total of about $85\pm10$ P donors in area
(V) from the leads and the secondary implant are present in the
Type~2 device with $w_s=l_s=100$~nm.

\begin{figure}[h!]
\includegraphics[width=6cm]{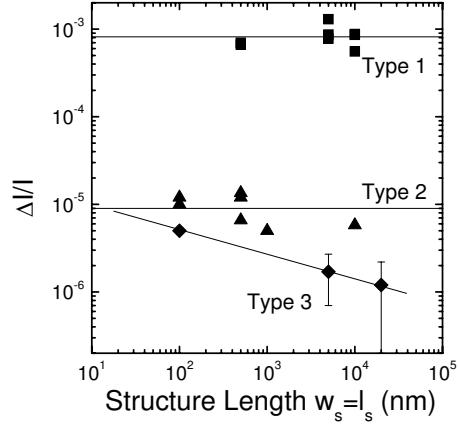} \caption{\label{fig:scaling}
Scaling of the EDMR signal intensity with varying structure size
$w_{s}$ for a series of devices of all three types investigated.
The lines are guides to the eye.}
\end{figure}

Figure~\ref{fig:scaling} shows the
EDMR signal intensity of the P hyperfine-split peak at
$B=348.88$~mT for all types of samples as a function of $w_s=l_s$.
Notably, the EDMR intensities of the Type~1 and 2 devices cluster
at $\Delta I/I\approx 10^{-3}$ and at $\Delta I/I\approx 10^{-5}$
respectively, independent of $w_s=l_s$. Due to Ohm's law and the
quadratic geometry of the gap, the resistance of each type of
device studied in a purely drift based model is independent of the
characteristic size and therefore the resonant changes in the
current should indeed be size independent. This demonstrates
that with the restriction of the current path to areas containing
few P donors, Pauli-blockade effects lead to an effective
influence of the transport properties of the device. Differences in
the EDMR signal intensities between the Types are expected due to
the different areas implanted with P and the current path not
being restricted completely to the gap area. Since we monitor
resonant changes in the photoconductivity, diffusive transport and
spin-dependent recombination throughout the whole sample cannot be
excluded outright. However, the fact that in the control
experiments on Type~3 devices, a significant spin-dependent
recombination is only observed for the smallest structure length,
where the relative contribution of the straggle to the P donors in
the active area (V) is largest, clearly indicates that the P
donors in the active area contribute most to the EDMR signal,
while the recombination near the leads or indiffused contacts only
plays a minor part.

Analyzing the signal-to-noise ratio $S/N$ of the spectra measured
and determining from them a signal-to-noise ratio $S/N|_1$ for a
single magnetic field scan based on Poisson statistics yields a
typical $S/N|_1\approx 5$ for Type~1 and $S/N|_1\approx 0.5$ for
Type~2 samples, again independent of $w_s=l_s$.  If the
independence of the signal intensity on the structure length
persist also at smaller length, single P donors could be monitored
in Type 2 samples with $w_s=l_s=14$~nm at the same $S/N$ ratio.
Nevertheless, the sensitivity demonstrated here on devices with as
few as 100 donors in the active area surpasses the sensitivity
demonstrated for EDMR so far by several orders of magnitude.

Remarkably, the relative signal intensities of the $P_{b0}$ and
the P hyperfine-split lines found in our experiments are of the
same order of magnitude, independent of the overall signal
intensity and the type of the device. This suggests that the
spin-dependent recombination process investigated is a P-$P_{b0}$
pair process, where a photogenerated excess electron is captured
by a P donor and forms a spin pair with a $P_{b0}$ defect, a
process also proposed as a readout scheme for silicon QIP
hardware\cite{boehme02}. Furthermore, P-$P_{b0}$ pair
recombination can only occur near the Si/SiO$_2$ interface, and
therefore any P donors in the bulk of the sample, underneath the
implanted leads and the indiffused contacts do not contribute to
the EDMR signal.

Whilst the results shown in Fig.~\ref{fig:scaling} indicate that
the signal is in fact due to the low-dose implant into the
metallic region, further investigations are warranted. Apart from
experiments investigating the influence of the defect density of
the Si/SiO$_2$ interface and the fabrication steps leading to the
$P_{b0}$ and dangling bond defects on the EDMR signals, devices
where the donor species used in the fabrication of the ohmic
contacts and leads is different from that in the active area
should be studied, so that any resonance signature detected can be
attributed to the active area unequivocally. Arsenic lends itself
to this purpose, as it has a different hyperfine splitting than
phosphorus. Initial measurements on devices with Arsenic leads and
ohmic contacts, both fabricated by ion implantation, show behaviour
consistent with the results presented here, which shows that the
signal is due to the donors implanted into the active region.


We have demonstrated a pathway to the spectroscopic study of a
small number of donors in semiconductors, by selective
implantation of phosphorus into silicon. We have shown that it is
possible to observe the change in conductivity caused by EDMR of
less than 100 donors in the active device region. In principle this
technique is not restricted to P, but can be extended to other dopants.


We would like to thank A. Ferguson, V. Chan, M. Stutzmann and A.
Stegner for helpful discussions and E. Gauja for technical
support. This work was supported by the Australian Research
Council, the Australian Government and by the US National Security
Agency (NSA), Advanced Research and Development Activity (ARDA),
the Army Research Office (ARO) under contract number
DAAD19-01-1-0653 and by the Deutsche Forschungsgemeinschaft (SFB
631).


\end{document}